\documentclass{elsart}
 \usepackage{graphicx}
 \usepackage{epsfig}

\usepackage{amssymb}

\begin{document}

\begin{frontmatter}

\title{Randall-Sundrum Reality at the LHC}

\author{Vernon Barger$^a$ and Muneyuki Ishida$^b$ }
\address{$^a$Department of Physics, University of Wisconsin, Madison, WI 53706, USA}
\address{$^b$Department of Physics, Meisei University, Hino, Tokyo 191-8506, Japan}

\begin{abstract}
The radion is expected to be the first signal of the Randall-Sundrum (RS) model. 
We explore the possibility of finding it in the ongoing Higgs searches at the LHC. 
The little RS model (LRS), which has a fundamental scale at $\sim 10^3$ TeV, 
is excluded over wide ranges of the radion mass from the latest $WW$ and $\gamma\gamma$ data by ATLAS and CMS.
\end{abstract}

\end{frontmatter}

\section{Introduction}

The Standard Model (SM) successfully explains almost all present experimental data; however,
it is an unsatisfactory model to be the ultimate theory of particle physics.
One of its defects is a large hierarchy between the two fundamental scales, 
the Planck scale $\bar M_{Pl}=M_{Pl}/\sqrt{8\pi}\simeq 2\times 10^{18}$~GeV and the weak scale $\sim 100$~GeV, which requires 
an unnatural fine-tuning of model-parameters when the model is applied to weak-scale phenomenology.
The Randall-Sundrum model\cite{RS} was originally proposed to solve this hierarchy problem.
RS introduced the five-dimensional anti-de Sitter spacetime, 
\begin{eqnarray}  
ds^2 &=& e^{-2ky}\eta_{\mu\nu}dx^\mu dx^\nu -dy^2\ ,
\label{eq1}
\end{eqnarray}
with a $S^1/Z_2$ compactified 5-th dimension, denoted as $y\in [0,L]$; 
there are two three-branes at $y=0$ and $y=L$, called UV and IR branes, respectively.
All the SM fields are confined to the IR brane in the original setup of Randall-Sundrum model, denoted as the RS1 model, and 
the 5-dimensional fundamental scale $M_5$ at UV brane is scaled down to $M_5e^{-kL}$ at the IR brane 
by the warp factor $e^{-kL}$ appearing in the metric of Eq.~(\ref{eq1}).
By taking $kL\simeq 35$, the fundamental scale $M_5=\bar M_{Pl}$ is scaled down to the TeV scale.
In order to suppress unwanted higher dimensional operators, which are not sufficiently suppressed by the TeV-scale cutoff 
on the IR brane in the RS1 model, 
SM gauge fields\cite{bulk,Pomarol,Rizzo,CHL} and fermions\cite{bulk10,bulk5} are considered to propagate in the bulk space. 
In this new setup\cite{bulk,Pomarol,Rizzo,CHL,bulk10,bulk5,bulk2,bulk3,bulk4,bulk6,bulk7,bulk8,bulk9,bulk11,bulk12}, denoted as the RS model, 
RS can address both the hierarchy problem and fermion mass hierarchies. 
From the electroweak precision measurements and various flavor physics,  
the new Kaluza-Klein (KK) modes of the bulk SM fields are constrained to be heavier than 
a few TeV\cite{bulk,Pomarol,Rizzo,CHL,bulk10,bulk5,bulk2,bulk3,bulk4,bulk6,bulk7,bulk8,bulk9,bulk11,bulk12,mass,mass2,mass3,mass4,mass5,mass6,CGHN}. 

The radion $\phi$ was introduced as a quantum fluctuation of the modulus $L$ of the 5-th dimension. 
It is necessary to stabilize $L$ to the above value. 
Goldberger and Wise showed that a bulk scalar field propagating in the background geometry, Eq.~(\ref{eq1}), 
can generate a potential that can stabilize $L$.\cite{GW} 
In order to reproduce the value $kL\simeq 35$,
the radion should have a lighter mass than that of the Kaluza-Klein modes of all bulk fields\cite{CGK}.
Thus, the detection of the radion may well be expected as the first signal to indicate that  
the RS model is truly realized in nature.

From a purely phenomenological perspective, the fundamental scale $M_5$ is scaled down far below 
the 4-dimensional $M_{Pl}$, by increasing the size $kL$ of the 5-th dimension. This is known as little Randall-Sundrum (LRS) model\cite{DPS}.
The hierarchy problem is yet unsolved in this model. 
The neutral kaon mixing parameter $\epsilon_K$ gives a strong constraint\cite{BCG} on $M_5$, namely $M_5>$ several thousand TeV, 
in the LRS model, and this corresponds to $kL\stackrel{>}{\scriptstyle \sim} 7$. 
The present precision measurements of SM flavors are consistent with $kL=7$.
The diphoton signal at the LHC is predicted to be largely enhanced\cite{DMS} in comparison to the RS model
with $kL=35$.

In this Letter we evaluate the production and decays of the radion.
Similar analyses were done in ref.\cite{CHL,DMS,GRW,Kingman,coll,coll2,coll3,Toharia2}.
We refine the calculations appropriate to the LHC experiments at 7 TeV (LHC7), 
and consider the possibility of finding $\phi$ in the ongoing LHC Higgs searches. 
We demonstrate that $\phi$ could be found in the SM Higgs search in 
the $\gamma\gamma$ and $W^*W$ channels at LHC7,
and consider the possibility of RS and LRS models being thereby tested.
We find that the LRS model with $kL=7$ is excluded by the 
latest ATLAS and CMS data over a wide range of the radion mass.

\section{Coupling to the SM particles}\ \ \ 
We define the fluctuation $F$ of the metric and the canonically normalized radion field $\phi$ as
\begin{eqnarray}
ds^2 &=& e^{-2(ky+F)}\eta_{\mu\nu}dx^\mu dx^\nu -(1+2F)^2 dy^2,\nonumber\\
F&=& \frac{\phi (x)}{\Lambda_\phi}e^{2k(y-L)}\ ,
\label{eq1-0}
\end{eqnarray}
where $\Lambda_\phi$ is the VEV of the radion field $\phi (x)$.

The couplings of the radion to the SM particles in the original RS model are composed of two parts,
\begin{eqnarray}
L &=& L_{trace} + L_{bulk}\ ,
\label{eq1-1}
\end{eqnarray}
where $L_{trace}$ is determined from general covariance\cite{GW,GRW,Kingman} to be
\begin{eqnarray}
L_{trace} &=& \frac{\phi}{\Lambda_\phi}T^\mu_\mu (SM).\nonumber\\
T^\mu_\mu (SM) &=& T^\mu_\mu (SM)^{\rm tree} + T^\mu_\mu (SM)^{\rm anom}\nonumber\\
  T^\mu_\mu(SM)^{\rm tree} &=& \sum_f m_f \bar ff - 2m_W^2 W_\mu^+ W^{-\ \mu} -m_Z^2 Z_\mu Z^\mu 
       + 2 m_h^2h^2-\partial_\mu h \partial^\mu h\nonumber\\
  T_\mu^\mu(SM)^{\rm anom} &=& -\frac{\alpha_s}{8\pi}b_{QCD}\sum_a F^a_{\mu\nu}F^{a\mu\nu}
       -\frac{\alpha}{8\pi}b_{EM} F_{\mu\nu}F^{\mu\nu}\ \ .
\label{eq1-2}
\end{eqnarray}
Here $T^\mu_\mu (SM)$, the trace of the SM energy-momentum tensor,
which is defined by $\sqrt{-g}T_{\mu\nu}(SM)=2\frac{\delta (\sqrt{-g}L_{SM})}{\delta g^{\mu\nu}}$, is represented as a sum of 
the tree-level term $T^\mu_\mu(SM)^{\rm tree}$ and the trace anomaly term $T_\mu^\mu(SM)^{\rm anom}$ for gluons and photons.
$F^a_{\mu\nu}(F_{\mu\nu})$ are their field strengths. 
The $b$ values are $b_{QCD}=11-(2/3)6+F_t$, including the top loop, and $b_{EM}=19/6-41/6 +(8/3)F_t-F_W$,
including the top and $W$ loops.\footnote{
$F_t=\tau_t(1+(1-\tau_t)f(\tau_t))$ and $F_W=2+3\tau_W+3\tau_W(2-\tau_W)f(\tau_W)$.
$f(\tau)=[{\rm Arcsin}\frac{1}{\sqrt{\tau}}]^2$ for $\tau\ge 1$ and $-\frac{1}{4}[{\rm ln}\frac{\eta_+}{\eta_-}-i\pi]^2$ for $\tau <1$
with $\eta_{\pm}=1\pm\sqrt{1-\tau}$ . 
Here $\tau_i\equiv \left(\frac{2m_i}{m_\phi}\right)^2$ for $i=t,W$.
}

The $T^\mu_\mu(SM)^{\rm tree}$ is proportional to particle masses.
The new RS model with the SM fields propagating in the bulk has an additional radion interaction, 
$L_{bulk}$, which is inversely proportional to the size of the 5-th dimension\cite{Rizzo,CHL}.
There is a correction to the interactions of fermions,  
massless gluons and photons that have couplings to a radion at tree level.

These interactions are very similar form to the interactions of SM Higgs
except for an overall proportional constant,\footnote{
The overall sign of the radion couplings is opposite to that of the Higgs couplings
in the most frequently used definition of $\phi (x)$, Eq.~(\ref{eq1-0}).
}
the inverse of the radion interaction scale $\Lambda_\phi$, which is the VEV of $\phi$.
It is given by
\begin{eqnarray}
\Lambda_\phi &=& e^{-kL}\sqrt{\frac{6 M_5{}^3}{k}}.
\label{eq1-3}
\end{eqnarray}
The five dimensional fundamental scale $M_5$ is related with $\bar M_{pl}$ by
\begin{eqnarray}
\bar M_{pl}^2 &=& \frac{M_5^3}{k}(1-e^{-2kL}) \simeq \frac{M_5^3}{k}\ .
\label{eq1-4}
\end{eqnarray}
Thus, Eq.~(\ref{eq1-3}) is rewritten by
\begin{eqnarray}
\Lambda_\phi &=& \sqrt{6}~ \tilde k  \frac{\bar M_{pl}}{k},\ \ \ \tilde k\equiv k e^{-kL}, 
\label{eqA}
\end{eqnarray}
where $\tilde k$ sets the mass scale of $KK$-excitations.

We adopt the radion effective interaction Lagrangian given in Ref.\cite{CHL}
The radion couplings to gluons and photons are
\begin{eqnarray}
L_A &=& -\frac{\phi}{4\Lambda_\phi}\left[\left( \frac{1}{kL} + \frac{\alpha_s}{2\pi}b_{QCD}\right) 
\sum_a F_{\mu\nu}^a F^{a\ \mu\nu} 
+\left( \frac{1}{kL} + \frac{\alpha}{2\pi}b_{EM} \right) 
F_{\mu\nu} F^{\mu\nu} \right]\ \ \ \ \ \ \ .
\label{eq2}
\end{eqnarray}
We note that $L_A$ has both contributions from $L_{bulk}$ proportional to $(kL)^{-1}$ 
and $L_{trace}$ from the trace anomaly term, 
while only the latter term contributes for the SM Higgs case.

The radion couplings to $W,Z$ bosons are
\begin{eqnarray}
L_V &=& -\frac{2\phi}{\Lambda_\phi}\left[ \left( \mu_W^2W_\mu^+W^{-\ \mu}+\frac{1}{4kL}W_{\mu\nu}^+W^{-\ \mu\nu}  \right)
+\left( \frac{\mu_Z^2}{2}Z_\mu Z^{\mu}+\frac{1}{8kL}Z_{\mu\nu} Z^{\mu\nu}  \right)\right] \ \ \ \ \ 
\label{eq3}
\end{eqnarray}
where $V_{\mu\nu}=\partial_\mu V_\nu -\partial_\nu V_\mu$ for $V_\mu=W_\mu^\pm ,Z_\mu$ and
$\mu_i^2$ ($i=W,Z$), which include the contributions from the bulk wave functions of $W,Z$, 
are represented by using $W(Z)$ mass $m_{W,Z}$ as
$\mu_i^2=m_i^2[1-\frac{kL}{2}(\frac{m_i}{\tilde k})^2]$.\footnote{
The physical masses of $W,Z$ bosons are identified with $\mu_i$, not $m_i\ (i=W,Z)$; however, 
we neglect small difference between $\mu_i$ and $m_i$, and the $m_i$'s are fixed with the physical masses.
}

The radion couplings to the fermions are proportional to their masses.
\begin{eqnarray}
L_f &=& -\frac{\phi}{\Lambda_\phi} m_f [I(c_L)+I(c_R)](\bar f_Lf_R + \bar f_Rf_L)\ .
\label{eq4}
\end{eqnarray}
The coupling is proportional to a factor $I(c_L)+I(c_R)$ that is dependent upon the bulk profile parameters
$c_L$ and $c_R$. The $I(c_L)+I(c_R)$ are given in ref.\cite{CHL} as $1\sim 1.19$ and 1 
for $b\bar b$ and $t\bar t,\tau\tau$ channels, respectively, 
while the value for $c\bar c$ is model-dependent. 
We simply take $I(c_L)+I(c_R)=1$ for all the relevant channels:
$b\bar b,t\bar t,\tau\tau,c\bar c$.\footnote{
$c\bar c$ has only a tiny $BF$. For $b\bar b$, $I(c_L)+I(c_R)$ is given as 1.66 in another example\cite{DPS}.
In that case  $BF(\phi\rightarrow bb)$ becomes about 2.5 times larger than our present result. 
}


The coupling of the radion to the IR brane-localized Higgs scalar $h$ is given by
\begin{eqnarray}
L_h &=& \frac{\phi}{\Lambda_\phi} (-\partial_\mu h\partial^\mu h + 2 m_h^2 h^2) \ .
\label{eq5}
\end{eqnarray}

The model parameters are $kL,\Lambda_\phi,m_\phi$ and $m_h$.
In the following we consider two values of $kL$: $kL=7$ corresponding to the LRS model and $kL=35$ to 
the original RS model. The value $\Lambda_\phi =3$~TeV is used\cite{DMS}.
$k$ is taken as $k<M_5$ in the original RS model\cite{RS}.
Here we simply take $k=M_5$. 
From Eq.~(\ref{eq1-3}) this corresponds to $\tilde k=\Lambda_\phi /\sqrt 6$.
The value of $m_h$ is taken as 130 GeV unless otherwise specified, while $m_\phi$ is treated as a free parameter.
By using the effective couplings Eqs.~(\ref{eq2}) - (\ref{eq5}) and these values of parameters, 
we calculate the partial decay widths and thier branching ratios in \S 4.

\section{Radion Production at the LHC}

\begin{figure}[htb]
\begin{center}
\resizebox{0.7\textwidth}{!}{
  \includegraphics{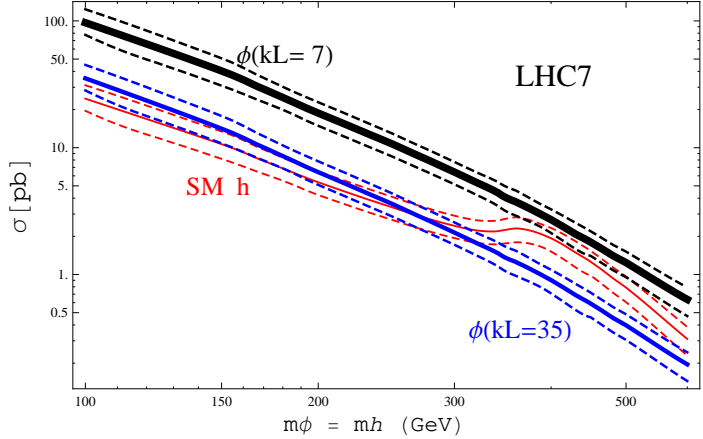}
}
\end{center}
\caption{The production cross sections at LHC7 of the radion $\phi$ via $gg$ fusion 
(solid blue $kL=35$ corresponding to the original RS model and solid thick black $kL=7$ to LRS model), 
compared with that of the SM Higgs with the same mass $m_{h^0}=m_{\phi}$(solid thin red).
The $\phi$ production cross sections are proportional to inverse-squared of $\Lambda_\phi$,
which is taken to be $\Lambda_\phi=3$~TeV here.
The overall theoretical uncertainties\cite{Djouadi} are denoted by dotted lines.
}
\label{fig1}
\end{figure}

The production cross section of the radion $\phi$ at hadron colliders
is expected to be mainly via $gg$ fusion, similarly to the production of a Higgs boson $h^0$.
These cross sections are proportional to the respective partial decay widths to $gg$.
The production cross section of $h^0$ has been calculated in NNLO\cite{Djouadi},
and by using this result\footnote{
The QCD radiative corrections to the tree level $gg\rightarrow h^0$ and $gg\rightarrow \phi$ should be equal
in the point-like approximation of the $gg\rightarrow h^0/\phi$ interactions, so 
we use the tree level result for $\Gamma (\phi\rightarrow gg)/\Gamma (h^0\rightarrow gg)$.
} 
we can directly estimate the production cross section of $\phi$ as
\begin{eqnarray}
\sigma(pp\rightarrow \phi X) &=& \sigma(pp\rightarrow h^0 X)\times
\frac{\Gamma(\phi\rightarrow gg)}{\Gamma(h^0\rightarrow gg)} \ .
\label{eq6}
\end{eqnarray}
By using the $\Gamma(\phi\rightarrow gg)$ partial width given later and 
$\Gamma(h^0\rightarrow gg)$ of the SM we can predict $\sigma(pp\rightarrow \phi X)$
in the two cases $kL=7,35$.
The result is compared with the SM Higgs production in Fig.~\ref{fig1}.

The production of $\phi$ scales with an overall factor $(1/\Lambda_\phi)^2$.
In the $\Lambda_\phi=3$~TeV case, the production of $\phi$ in the original RS model ($kL=35$)
is almost the same as that of the SM Higgs boson of the same mass, while in the LRS case (kL=7) 
the radion cross section exceeds that of the SM Higgs.
This is because $\phi$ production from $gg$ fusion has an amplitude that includes a term
proportional to $1/kL$ at tree level, while there is no such term in $h^0$ production. 
Our prediction of $\sigma(\phi)$ in Fig.\ref{fig1} 
includes the $\pm 25$\% 
uncertainty associated with the theoretical uncertainty on $\sigma(h^0)$\cite{Djouadi}.

\section{Radion Decay}\ \ \ \ 
For the radion decay channels $\phi\rightarrow AB$,
we consider $AB=gg,\gamma\gamma$, $W^+W^-$, $ZZ,b\bar b,c\bar c,\tau^+\tau^-,t\bar t$, and $h^0h^0$.
Their partial widths are given by the formula
\begin{eqnarray}
\Gamma (\phi \rightarrow AB) &=& 
\frac{N_c}{8\pi (1+\delta_{AB})m_\phi^2} p(m_{\phi}^2;m_A^2,m_B^2)\times |M|^2
\label{eq7}
\end{eqnarray} 
where $p$ is the momentum of particle $A$(or $B$) in the CM system
and $|M|^2$ represents the decay amplitude squared which are given in 
Table~\ref{tab1}.\footnote{
For the $gg$ decay of $\phi$ we include the radiative corrections at NNLO by using the K-factor from 
Higgs production.
 We use central values of the K-factor given in Fig. 8 of 
 Ref.\cite{Kfact}: For $m_{h^0}=100\sim 600$~GeV, $K=2.0\sim 2.6$.
 This value is about 10\% larger than the K-factor of Higgs decaying into $gg$ in NNLO given in ref.\cite{decayK} 
 but within the uncertainty of the choice of the renormalization scale. So we assume the $\phi$ and $h$ $K$-factors 
are equal and adopt the value in ref.\cite{Kfact} .
}

\begin{table}[h]
\caption{Decay amplitudes squared $|M|^2$ of $\phi$ decays. 
$|M|^2$ for $ZZ$ is obtained by replacement $W\rightarrow Z$ from $M_{T,L}^{WW}$. 
$gg$ includes a K-factor 
$K(m_\phi)$\cite{Djouadi}.
$b_{QCD},b_{EM}$ are given below Eq.~(\ref{eq1-2}) in the text.
$b\bar b$ includes radiative corrections\cite{Hunter} 
of running mass $m_b(m_\phi)$ and an overall factor $C(m_\phi)$,
but we adopt a fixed mass $m_b=4.5$~GeV for the kinematical factor $(m_\phi^2/4-m_b^2)^{3/2}$. 
Similar expressions are also applied to $t\bar t,c\bar c$ channels.  
The off-shell $WW^*(ZZ^*)$ channels in the low-mass $\phi$ case are treated 
in the same method as in ref.\cite{Keung}.
}
\begin{center}
\begin{tabular}{l|l@{}l}
decay channel & $|M|^2$ &  \\
\hline
\begin{tabular}{l} $W^+W^-$\\ $gg$\\ $\gamma\gamma$ \end{tabular} 
 & \begin{tabular}{l} $2|M_T^{WW}|^2+|M_L^{WW}|^2$\\  $2|M_T^{gg}|^2\cdot K(m_\phi)$\\ $2|M_T^{\gamma\gamma}|^2$ \end{tabular} 
 & \\
\ \ $b\bar b$ & \multicolumn{2}{l}{$\frac{8}{\Lambda_\phi^2}\ C(m_\phi)\  m_b(m_\phi)^2\ (\frac{m_\phi^2}{4}-m_b^2)$} \\
\ \ $\tau^+\tau^-$ & \multicolumn{2}{l}{$\frac{8}{\Lambda_\phi^2} m_\tau^2 (m_\phi^2/4-m_\tau^2)$} \\
\ \ $h^0h^0$ & \multicolumn{2}{l}{$|-\frac{1}{\Lambda_\phi}(m_\phi^2+2m_{h^0}^2)|^2$} \\ 
\multicolumn{3}{l}{
\begin{tabular}{|l|}\hline $M_T^{WW}=-\frac{2}{\Lambda_\phi}\{  \mu_W^2 -\frac{1}{2kL}\frac{m_\phi^2-2m_W^2}{2}  \}$\\
       $M_L^{WW}=-(1-\frac{m_\phi^2}{2m_W^2})M_T^{WW}-\frac{2}{\Lambda_\phi}\frac{1}{2kL}\frac{m_\phi^2(m_\phi^2/4-m_W^2)}{m_W^2}$\\
       $M_T^{gg,\gamma\gamma}=\frac{m_\phi^2}{2\Lambda_\phi kL}(1+\frac{\alpha_s\ b_{QCD}}{2\pi}  kL)$,\ \ 
       $\frac{m_\phi^2}{2\Lambda_\phi kL}(1+\frac{\alpha\ b_{EM}}{2\pi}  kL)$\\
     \hline
   \end{tabular}}\\
\end{tabular}
\end{center}
\label{tab1}
\end{table}

\begin{figure}[htb]
\begin{center}
\resizebox{0.8\textwidth}{!}{
  \includegraphics{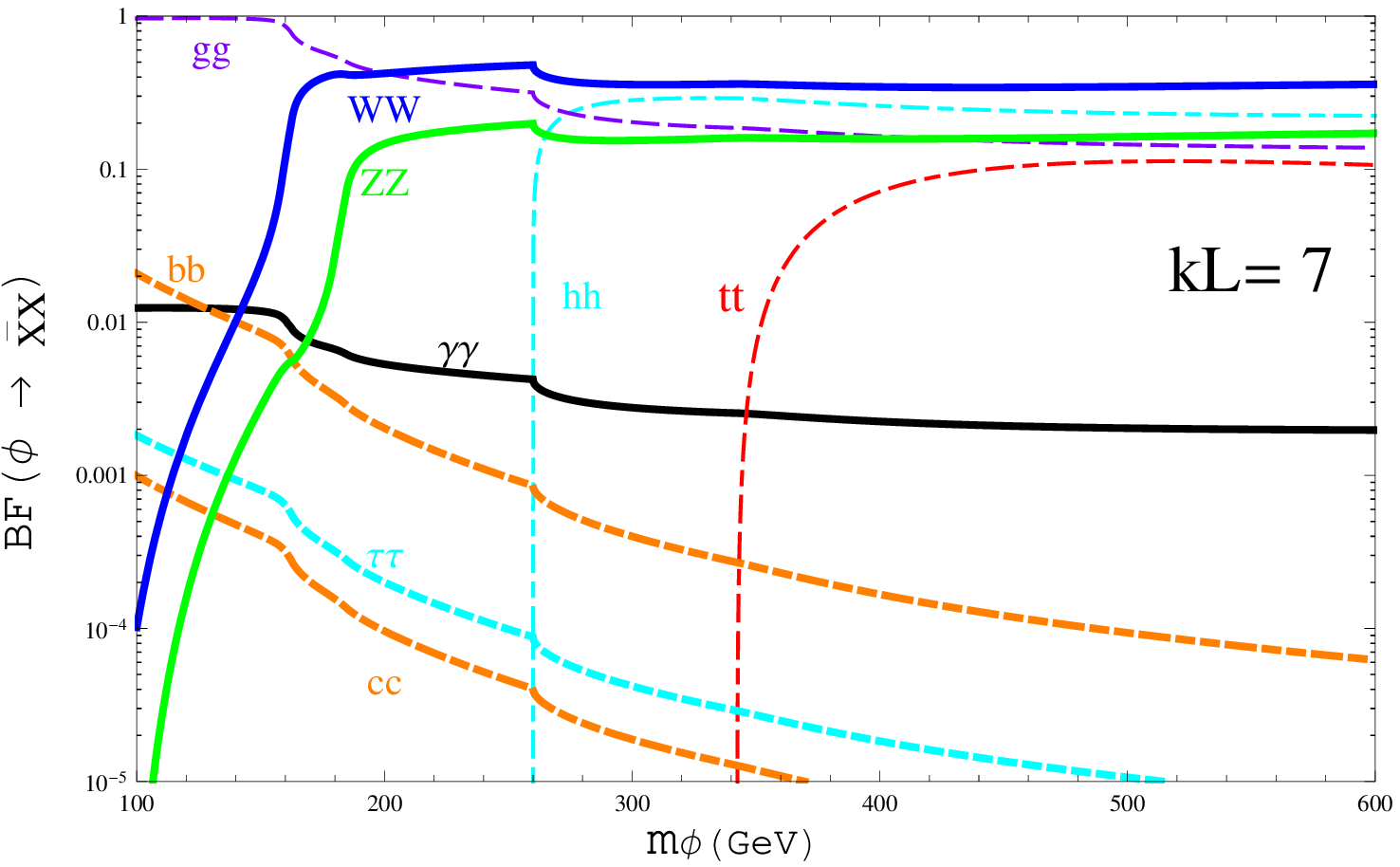}
}
\resizebox{0.8\textwidth}{!}{
  \includegraphics{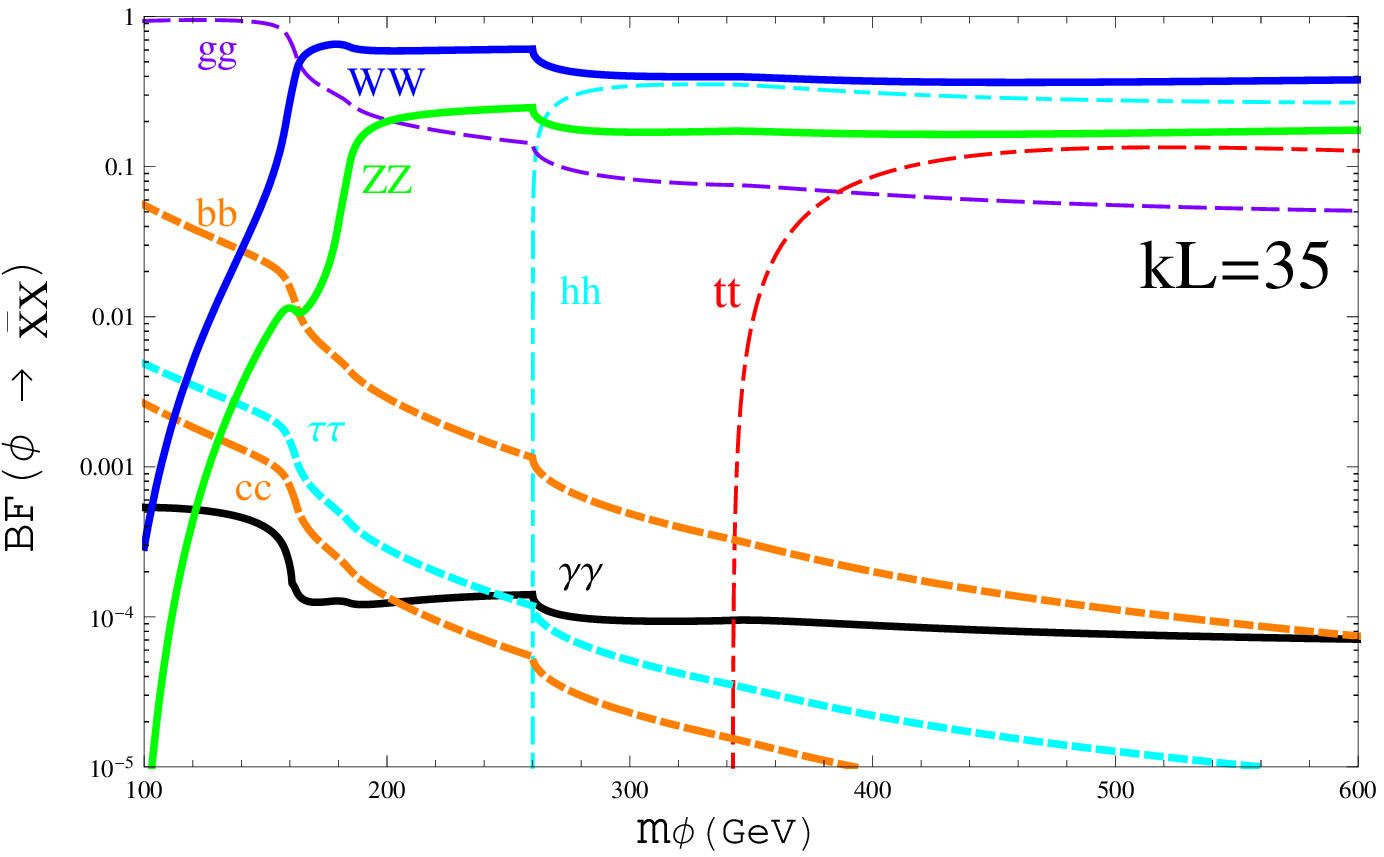}
}
\end{center}
\caption{Decay Branching Fractions of $\phi$ versus $m_{\phi}$(GeV) 
for $kL =7$(LRS model) and 35(RS model). $m_{h^0}$ is taken to be 130 GeV. 
}
\label{fig2}
\end{figure}

The results of the decay branching fractions are given for the two cases $kL=7,35$ in Fig.~\ref{fig2}.

In the $kL=7$(LRS model) case, there is a strong enhancement of $BF(\phi\rightarrow\gamma\gamma)$,
in comparison to the RS model
with $kL=35$, as was pointed out in ref.\cite{DMS}. 
The  BF($\gamma\gamma$) reaches almost $10^{-2}$ in LRS model, 
while it is $\sim 10^{-4}$ almost independent of $m_\phi$ in RS model. 
BF$(\gamma\gamma)$ is proportional to $(1/kL)^2$ 
in the $m_\phi \stackrel{>}{\scriptscriptstyle \sim} 200$~GeV region
in the LRS model.\footnote{
It should be noted that in the original RS model setup where the SM fields are confined in the IR brane,
BF($\gamma\gamma$) steeply decreases with $m_\phi > 200$ GeV similarly to the SM Higgs. 
This modified behavior comes from $L_{bulk}$
in the new bulk field scenario of SM field.}
This huge enhancement is from the bulk field coupling of $\phi$ proportional to $1/kL$.  
We do not find sharp dips in $BF(\phi\rightarrow WW,ZZ)$ around $m_\phi\simeq 450$ GeV of Fig.~1 
in ref.\cite{DMS}.

\begin{figure}[htb]
\begin{center}
\resizebox{0.8\textwidth}{!}{
  \includegraphics{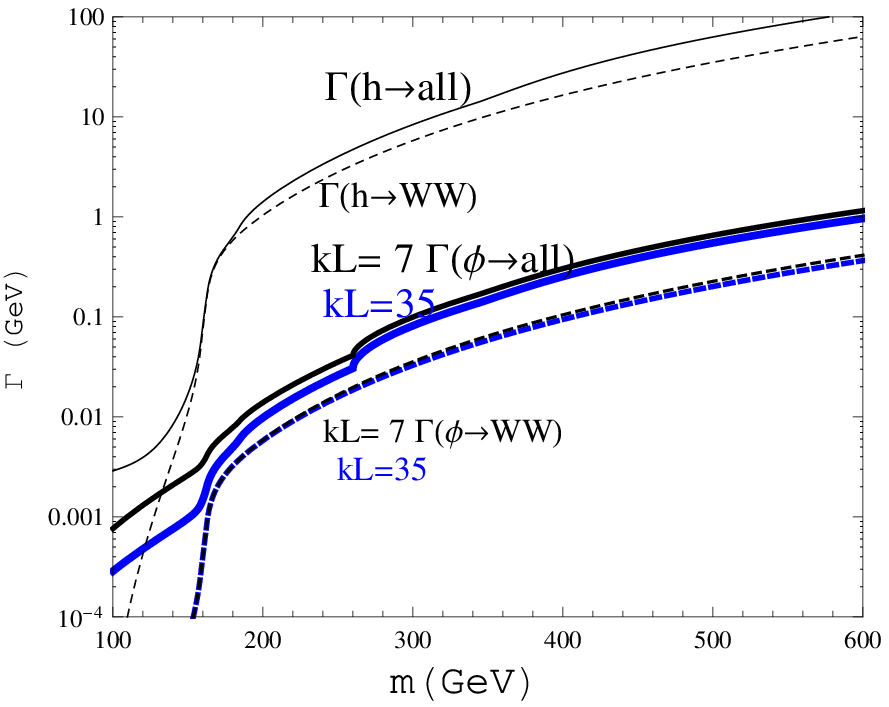}
}
\end{center}
\caption{Total widths (GeV) and $W^+W^-$ partial widths of $\phi$ compared with the SM higgs $h^0$ with the same mass $m=m_\phi=m_h$(GeV). 
The $\Lambda_\phi =3$~TeV and $kL=7,35$ cases are shown. For $\Gamma (\phi\rightarrow {\rm all})$ the $m_h$ is fixed with 130 GeV.
 The widths of $\phi$ are proportional to the inverse squared
of $\Lambda_\phi$. 
}
\label{fig3}
\end{figure}

The total width of $\phi$ in Fig.\ref{fig3} is one or two orders of magnitude smaller 
than that of the SM Higgs of the same mass.
This is because the choice of $\Lambda_\phi = 3$~TeV is about one order of magnitude larger than the Higgs VEV $v=246$~GeV.
A $\phi$ resonance would be observed with the width of the experimental resolution.
The $\Gamma (\phi\rightarrow WW)$ partial width is negligibly small compared with 
 $\Gamma (h^0\rightarrow WW)$, and thus $\phi$ production via vector boson($WW,ZZ$) fusion
is unimportant at the LHC, providing another way to distinguish $\phi$ and $h^0$.
We note that double Higgs production via $\phi$ decays would uniquely distinguish $\phi$ and $h$.

\begin{figure}[htb]
\begin{center}
\resizebox{0.8\textwidth}{!}{
  \includegraphics{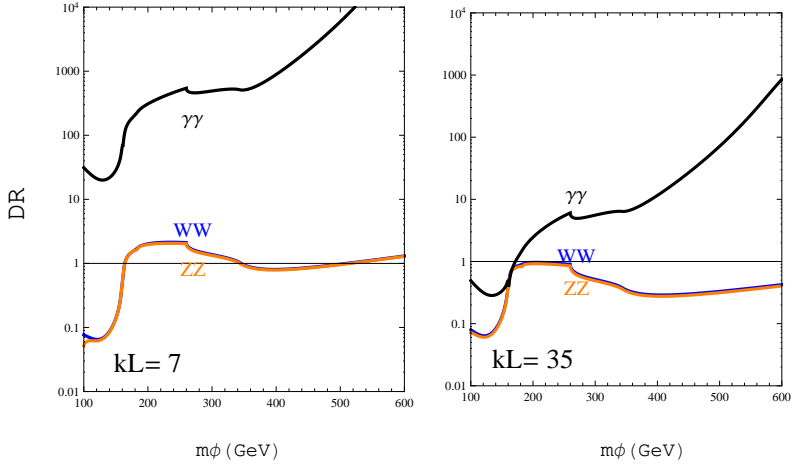}
}
\end{center}
\caption{$\phi$ Detection Ratio ($DR$) to the SM higgs $h^0$  of Eq.~(\ref{eq8}) 
for the $\bar XX=W^+W^-$(solid blue), $ZZ$(dashed orange), and $\gamma\gamma$(solid black)
final states for $kL =7,35$ versus $m_{\phi}$(GeV).
}
\label{fig4}
\end{figure}

\section{Radion Detection compared to SM Higgs}
Next we consider the detection of $\phi$ via the $W^+W^-$, $ZZ$ and $\gamma\gamma$ decay channels.
The $\phi$ search can be made in conjunction with the Higgs search.
The properties of $h^0$  at the LHC are well known, so we use them 
as benchmarks of the search for $\phi$.  

The $\phi$ detection ratio ($DR$) to $h^0$ in the $\bar XX$ channel is
defined\cite{book} by
\begin{eqnarray}
DR & \equiv &
\frac{\displaystyle \Gamma_{\phi\rightarrow gg}\Gamma_{\phi\rightarrow \bar XX}/
\Gamma_{\phi}^{\rm tot}}{
\displaystyle \Gamma_{h^0\rightarrow gg}\Gamma_{h^0\rightarrow \bar XX}/\Gamma_{h^0}^{\rm tot}}\ ,\ \ \ \ \ \ \ \ \ 
\label{eq8} 
\end{eqnarray}
where $\bar XX=W^+W^-,\ ZZ,$ and $\gamma\gamma$. 
The $DR$ are plotted versus $m_{\phi}=m_{h^0}$ in Fig.~\ref{fig4} for the two cases $kL=7$ and 35.

\begin{figure}[htb]
\begin{center}
\resizebox{0.6\textwidth}{!}{
  \includegraphics{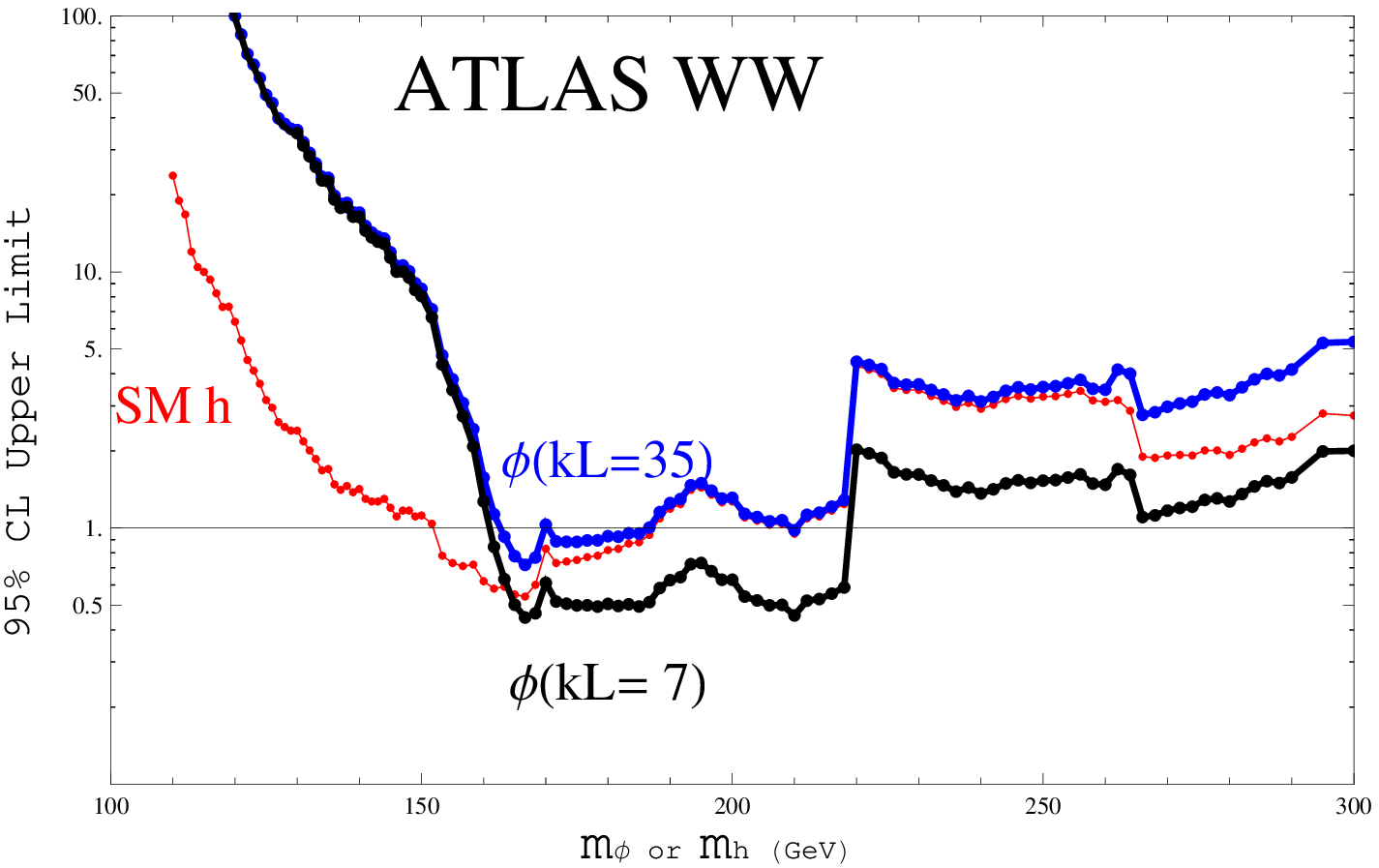}
}
\resizebox{0.6\textwidth}{!}{
  \includegraphics{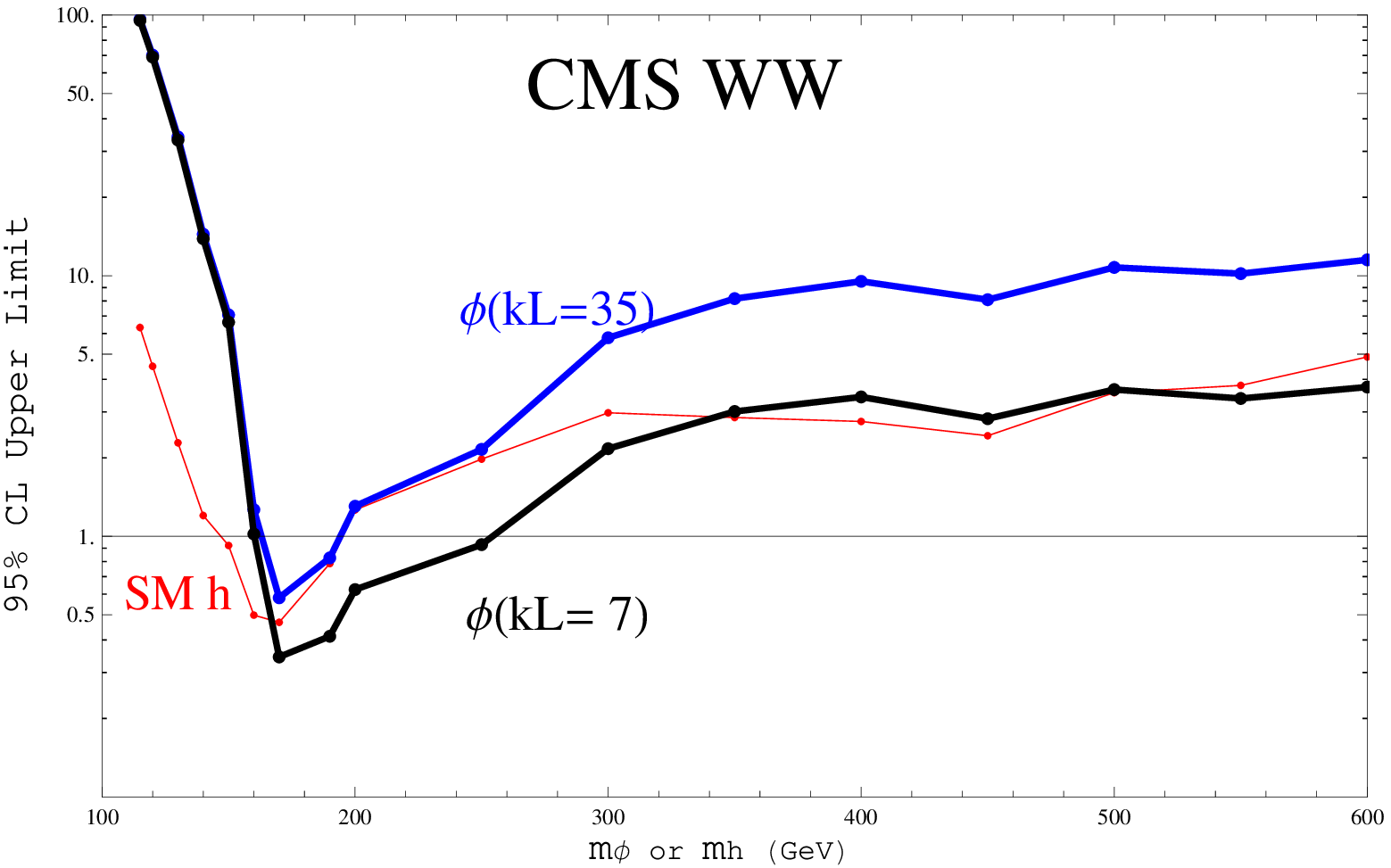}
}
\resizebox{0.6\textwidth}{!}{
  \includegraphics{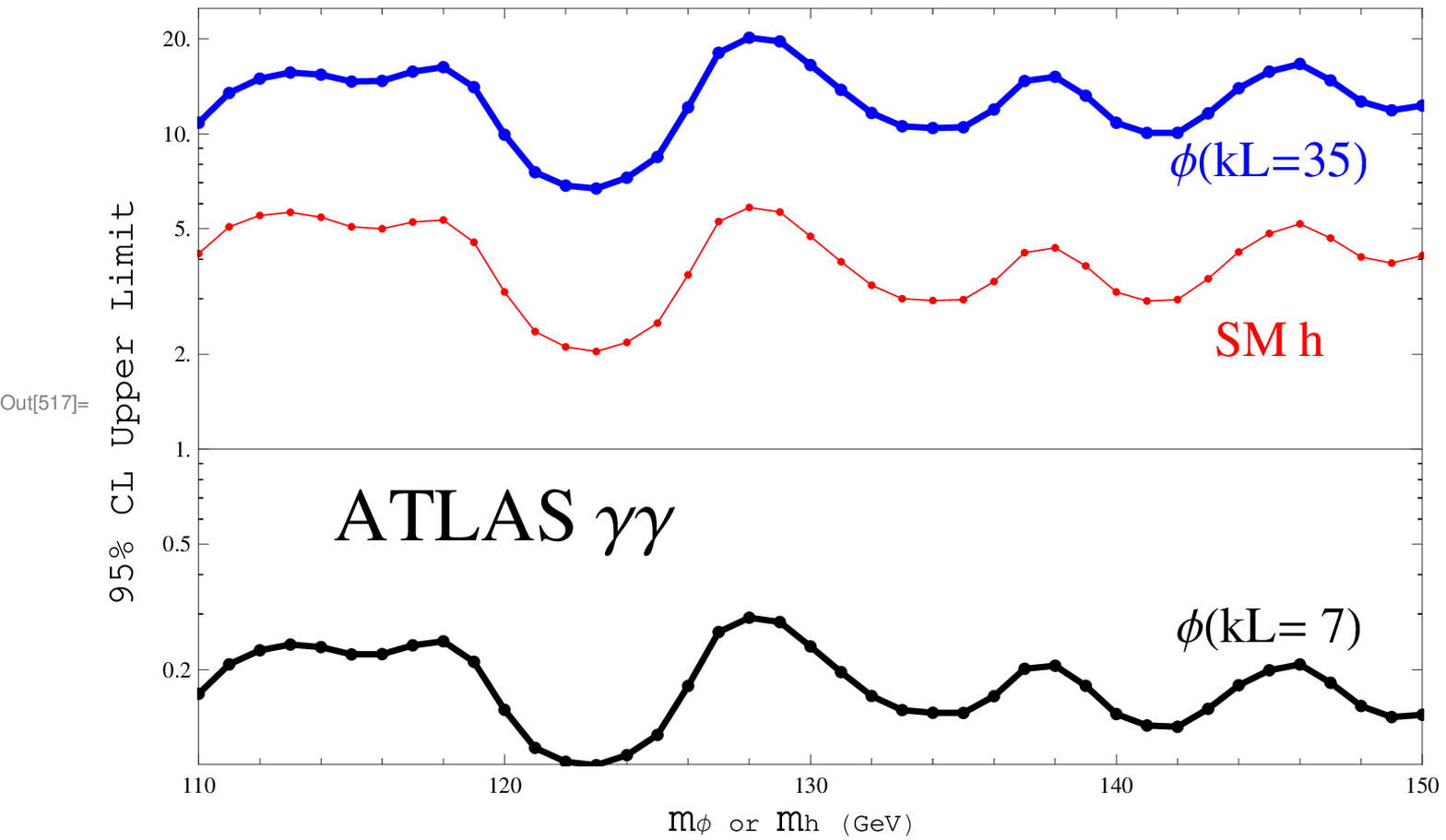}
}
\end{center}
\caption{The 95\% confidence level upper limits on
$(1/DR)\times (\sigma_{\rm exp}/\sigma(h^0\rightarrow \bar XX))$.
This is the signal of a scalar boson decaying into $\bar XX$ relative to the radion cross section 
$[\sigma(\phi \rightarrow \bar XX)=\sigma(h^0 \rightarrow \bar XX)\times DR ]$
for $\bar XX=WW\rightarrow l\nu l\nu$(ATLAS\cite{AtlasWW}, CMS\cite{cmsWW}) and $\gamma\gamma$
(ATLAS\cite{Atlasll}) data.  
The cases $kL=35$(solid blue) and 7(solid black) are shown.
Similar results for the SM Higgs boson are also given(red thin-solid curve).  }
\label{fig5}
\end{figure}

In mass range between the $WW$ threshold and the $h^0h^0$ threshold (300 GeV in the present illustration)
the $\phi$ to $h^0$ detection ratio is relatively large in both $WW$ and $ZZ$ channels.
The $DR$ is almost 2 in the $kL=7$ case. 
The $DR$ in $\gamma\gamma$ channel increases rapidly in the large $m_\phi=m_{h^0}$ region 
since $\Gamma_{h^0\rightarrow\gamma\gamma}$ steeply decreases with increasing $m_h$
due to the cancellation between top and $W$ loops.
So the $\gamma\gamma$ channel used in the search for the SH Higgs in the mass range 115-150 GeV 
by the LHC experiments is more sensitive for the $\phi$ search.
Surprisingly large enhancements of $DR$ in the $\gamma\gamma$ channel are predicted in this mass region in the $kL=7$ case.
This is because the $BF(\phi\rightarrow\gamma\gamma)$ is hugely enhanced in LRS model, as explained in the previous section. 
The $\phi$ should be detected in $\gamma\gamma$ in the current LHC data in the LRS scenario.
This possibility is checked in Fig.~\ref{fig5}.

The cross-section of a putative Higgs-boson signal,
relative to the Standard Model cross section, as a function of the assumed Higgs boson mass, 
is widely used by the experimental groups to determine the allowed and excluded regions of $m_{h^0}$.
By use of the $DR$ in Fig.~\ref{fig4}, we can determine the allowed region of
$m_{\phi}$ from the present LHC data.
Figure~\ref{fig5} shows the 95\% confidence level upper limits on Higgs-like $\phi$ signals
decaying into $\bar XX$ versus $m_{\phi}$ for $\bar XX=WW$(ATLAS\cite{AtlasWW},CMS\cite{cmsWW}) and 
$\gamma\gamma$(ATLAS\cite{Atlasll}). 

For the LRS model with $kL=7$, $m_{\phi}$ is excluded by ATLAS data at 95\% CL over the $m_{\phi}$ range,
$160< m_\phi < 220$~GeV,
while for RS model $kL =35$ almost no regions of $m_{\phi}$ are yet excluded. 
Similar results are found from CMS data\cite{cmsWW}.

The $\phi$ search is also applicable to the Tevatron data.
The CDF and D0 experiments excluded a SM Higgs with mass $158\ {\rm GeV}< m_{h^0} < 175$~GeV 
from data on $WW,ZZ$ channels.
The same data exclude $\phi$ in the $kL=7$ case in $m_\phi$ range,
$163\ {\rm GeV} < m_{\phi} < 180$~GeV, while 
for $kL=35$, only $165\ {\rm GeV} < m_{\phi} < 171$~GeV is excluded.

The $\gamma\gamma$ final state is very promising for $\phi$ detection, because
the $\phi$ to $h^0$ detection ratio 
is generally very large in all the mass range of $m_{\phi}$, as shown in Fig.~\ref{fig4}.
The $\gamma\gamma$ data of ATLAS do not show any resonance enhancements in $110<m<150$~GeV
($cf.$  Fig.~\ref{fig5}, so $\phi$ is excluded in this mass region in the LRS model ($kL=7$) case,
while no $m_\phi$ regions are excluded in the RS model with $kL=35$. 
The radion upper limits in the diphoton channel have similar shapes of the curve for the Higgs
because the detection ratios are relatively constant in this narrow mass range
$m=110\sim 150$~GeV: See Fig.~4.  

For $m_{\phi} > 150$~GeV, the $\gamma\gamma$ signal of $h^0$ is too small
to be detected at present, but future data in this region can determine the existence of $\phi$.

We take $\Lambda_\phi=3$~TeV in our analyses.
By taking larger values of $\Lambda_\phi$, the detection ratios of $\phi$ decrease since
the $\phi$ production cross section is proportional to $(1/\Lambda_\phi)^2$.
By taking $\Lambda_\phi >5$~TeV, all values of $m_\phi$ become allowed by the latest ATLAS and CMS  $WW^*$ data, 
By taking $\Lambda_\phi >10$~TeV, all values of $m_\phi$ become allowed by the latest ATLAS $\gamma\gamma$ data, 

A comment should be added here. There is a possible mixing effect\cite{GRW} between the radion and SM Higgs boson
through the action
\begin{eqnarray}
S_\xi &=& -\xi\int d^4x \sqrt{-g}R\ H^\dagger H
\label{eq14}
\end{eqnarray}
where $H$ is Higgs doublet and $H=((v+h^0)/\sqrt 2,0)$.
Because of the Higgs-like nature of the radion coupling, its effect can be very large even if the mixing angle is very small. 
We have excluded wide $m_\phi$-regions in LRS model with $\Lambda_\phi=3$~TeV in the no-mixing case. 
The effect of Eq.~(\ref{eq14}) is studied in detail in ref.\cite{Toharia} including a large $\xi$ case of the RS1 model\cite{coll2}.
Generally speaking, when BF($\phi\rightarrow WW/ZZ$) becomes larger(smaller) owing to the mixing effect,
BF($\phi\rightarrow \gamma\gamma$) becomes smaller(larger) than in the no-mixing case. 
So the $WW/ZZ$ and $\gamma\gamma$ channels are complementary for the detection of the radion.

\section{Concluding Remarks}
We have investigated the possibility of finding the radion $\phi$ at the Tevatron and LHC7.
The radion can be discovered in the $WW$, $ZZ$, and $\gamma\gamma$ channels in the search for
the SM Higgs $h^0$. 
The $WW$ signal rate can be comparable to that of the SM $h^0$,
in the mass region $m_{\phi}\sim 160~{\rm GeV}$ up to $2m_{h^0}$ as shown in Fig.~\ref{fig4}. 
The $\gamma\gamma$ search channel is especially promising,
since $BF(\phi\rightarrow\gamma\gamma)$ is almost constant 
at $\sim 10^{-4}$ for $m_\phi$ below 600~GeV, and
the corresponding $\phi$ detection ratio compared to $h^0$ is very large above $m_\phi=180$~GeV.
Combining the $WW,\gamma\gamma$ data of ATLAS and CMS,
the LRS model with $kL=7$ is already excluded over wide ranges of $m_\phi$ for $\Lambda_\phi=3$~TeV.

\noindent{\it Note added}\ \ \ \ 
A constraint on the mass of the first KK-graviton, $m_{G1}$,
was reported\cite{mGATLAS} by the ATLAS collaboration from the 
2-photon channel: $m_{G1}>0.80(1.95)$~TeV 
for $k/\bar M_{Pl}=0.01(0.1)$. $m_{G1}$ is given by $x_1\tilde k$, 
where $x_1=3.83$ is the first root of $J_1$ bessel function, and thus,
a very strong constraint is obtained for the $\Lambda_\phi$ in Eq.~(\ref{eqA}): 
$\Lambda_\phi = \sqrt{6}\frac{m_{G1}}{x_1}\left(\frac{k}{\bar M_{pl}}\right)^{-1}$ which gives
$\Lambda_\phi > 50(12)$~TeV for $\frac{k}{\bar M_{Pl}}=0.01(0.1)$. 
However, the lower limit of $\Lambda_\phi$ is strongly dependent upon the 
value of tha ratio $\frac{k}{\bar M_{Pl}}$, and if this ratio is taken to be unity\cite{mGHooman},
the lower limit on $\Lambda_\phi$ will be relaxed to a few TeV; 
The curvature $k$ should be less than the Planck mass $\bar M_{Pl}$ in the RS model\cite{RS}.

\noindent\underline{\it Acknowledgements}

The authors would like to express their sincere thanks to Prof. W.-Y. Keung for his collaboration.
M.I. is very grateful to the members of phenomenology institute of University of Wisconsin-Madison for hospitalities.
This work was supported in part by the U.S. Department of Energy under grant No. DE-FG02-95ER40896,
in part by KAKENHI(2274015, Grant-in-Aid for Young Scientists(B)) and in part by grant
as Special Researcher of Meisei University.

\nocite{*}

\bibliography{apssamp}

\end{document}